\documentstyle[preprint,prabib,aps,12pt]{revtex}

\begin{document}
\title{Flux Tubes in Weyl Gravity}
\author{V. Dzhunushaliev
\thanks{E-Mail Addresses : dzhun@rz.uni-potsdam.de and 
dzhun@freenet.bishkek.su; \qquad permanent address: 
Dept. Theor. Phys., Kyrgyz State National University, 
Bishkek 720024, Kyrgyzstan}}
\address{Institut f\"ur Mathematik, Universit\"at Potsdam 
PF 601553, D-14415 Potsdam, Germany} 
\author{H.-J. Schmidt
\thanks{http://www.physik.fu-berlin.de/\~{}hjschmi \ \ 
 \quad  hjschmi@rz.uni-potsdam.de}}
\address{Institut f\"ur Theoretische Physik, Freie
Universit\"at Berlin\\
and \\
Institut f\"ur Mathematik, Universit\"at Potsdam 
PF 601553, D-14415 Potsdam, Germany}

\maketitle

\begin{abstract}
The spherically symmetric solutions in  Weyl 
gravity interacting with $U(1)$ or $SU(2)$ 
 gauge fields are examined. It is shown that 
these solutions are conformally 
equivalent to an infinite flux tube with  constant 
(color) electric  and 
magnetic fields. This allows us to say that  Weyl 
gravity has in some sense a 
classical confinement mechanism. We discuss a possible connection 
with flux tubes in quantum chromodynamics.
 \end{abstract}

\section{Introduction}

The conformal Weyl gravity has a long history. Initially 
it was introduced as 
 an attempt of generalizing  Einstein gravity 
to obtain a common description for the unified theory of 
gravity+electromagnetism, see \cite{xxx}. This attempt was 
unsuccessful,
however, 
 now this theory attracts much attention of  gravity 
researchers, see \cite{vdschmidt1} and the references cited there. 

In this paper we want to demonstrate an 
unexpected property of this theory of gravity as a classical 
 confinement 
for $U(1)$ and $SU(2)$ gauge fields. 
\par 
What is it a confinement in QCD? There is an assumption 
that the force lines of color-electrical field between 
quark and antiquark are confined to a flux tube stretched between 
these particles. This object is a complete quantum phenomenon. 
The difficulties to explaining  this tube is that the $SU(3)$ 
nonabelian gauge field is a strongly  nonlinear field,
 and we cannot use the Feynman diagramm technique. 
\par 
In this paper we will show that spherically symmetric solutions 
in  Weyl gravity model  in the some sense
 properties of 
 the classical confinement 
mechanism. 

\section{(2+2)-decomposition of the  metric}

Here we examine the static spherically symmetric solutions in 
 four-dimensional 
(4D) Weyl gravity with $U(1)$ and $SU(2)$ gauge fields. The 
most general 
spherically symmetric metric takes the form: 
\begin{equation}
ds^2 = g_{AB}dx^Adx^B  - a^2(r) d\Omega ^2 ,
\label{b-1}
\end{equation}
where $g_{AB}$ depends on the coordinates $x^A$ only, has 
curvature scalar 
$X$ and signature $(+-)$, ($A,B = 0,1$, $x^0=t, \ x^1=r$, 
but $r$
 need not be space-like) and 
$d\Omega ^2 = d\theta ^2 + \sin ^2 \theta d \varphi ^2$ 
is the metric on the unit $S^2$ sphere. As the Weyl gravity, 
Maxwell electrodynamic and Yang-Mills theory are  conformally 
invariant theories we can work 
with the following metric which is conformally equivalent to 
(\ref{b-1}) metric: 
\begin{equation}
ds^2 = \tilde g_{AB} dx^A dx^B - d\Omega ^2 ,
\label{b-2}
\end{equation}
where $\tilde g_{AB} = g_{AB}/a^2$. The hypersurface $t=const$ 
for this metric is $R \times S^2$, i.e. a tube 
with the constant cross section as the $S^2$ sphere. 
\par 
Now we repeat a (2+2)-decomposition technique following to paper 
\cite{vdschmidt1}. 
\par 
Let the 4D metric be 
\begin{equation}
ds^2  =   g_{ij} dx^i dx^j , \qquad i,j=0, \dots 3 .
\label{a-1}
\end{equation}
Let we have the following ansatz:
\begin{equation}
ds^2  =   d\sigma^2 - d\tau^2
\label{a-2}
\end{equation}
where $d\sigma^2$ and $d\tau^2$  are both 2-dimensional. The 
metric 
\begin{equation}
d\sigma^2   = \tilde  g_{AB} dx^A dx^B  , \qquad A,B=0,1  .
\label{a-4}
\end{equation}
The other 2-dimensional metric 
\begin{equation}
d\tau^2   =   g_{\alpha \beta} dx^{\alpha} dx^{\beta} , \qquad
 \alpha, \beta = 2,3 ,
\label{a-5}
\end{equation}
where $g_{\alpha \beta}$ depends on the $x^{\alpha}$ only, has 
curvature
scalar $Y$ and signature $(++)$.\footnote{For us is only important 
that  
$d\tau ^2 = d\theta ^2 + \sin ^2 \theta d\varphi ^2$ with $Y = 2$.}
\par 
The lagrangian for the conformal Weyl gravity interacting 
with the gauge field is: 
\begin{equation}
{\cal L} = -C_{ijkl} C^{ijkl} - \frac{\kappa}{4} tr
\left (F_{lm}F^{lm} \right ) ,
\label{a-6}
\end{equation}
here $C_{ijkl}$ is the conformally invariant Weyl tensor, 
$F_{lm}$ is the tensor of field  strength  for the gauge potential 
and 
we keep both
 possible signs of $\kappa$. 
The Bach equations are: 
\begin{equation}
B_{ij} = \kappa T_{ij} ,
\label{a-7}
\end{equation}
where 
\begin{eqnarray}
B_{ij} =  B^{(1)}_{ij}  + B^{(2)}_{ij}   
\label{a-8}\\
B^{(1)}_{ij} =
- \Box R_{ij} + 2 R^{\ k}_{i \ ;jk} - \frac{2}{3} R_{;ij}
 + \frac{1}{6} g_{ij} \Box R
\label{a-9}\\
B^{(2)}_{ij}  \ = \ \frac{2}{3} R \ R_{ij}
- 2 R_{ik}R^k_j - \frac{1}{6} R^2 g_{ij}
+ \frac{1}{2} g_{ij} R_{kl} R^{kl} .
\label{a-10}
\end{eqnarray}
It can be shown that for our (\ref{a-1}), (\ref{a-4}), (\ref{a-5}) 
ansatz:
\begin{eqnarray}
B_{AB} = 
\frac{1}{3}X_{;AB} + g_{AB}
\left(\frac{1}{6}\Box Y - \frac{1}{3} \Box X + \frac{1}{12} 
Y^2 - \frac{1}{12} X^2 \right) & = & \kappa T_{AB} ,
\label{a-11}\\ 
B_{\alpha \beta} = 
\frac{1}{3}Y_{;\alpha \beta} + g_{\alpha \beta}
\left(\frac{1}{6}\Box X - \frac{1}{3} \Box Y + \frac{1}{12} 
X^2 - \frac{1}{12}Y^2 \right) & = & \kappa T_{\alpha \beta} .
\label{a-12} 
\end{eqnarray}

\section{$U(1)$ flux tube}

In Ref.\cite{riegert84}  the static, spherically symmetric solution 
for the Bach-Maxwell equations is given: 
\begin{equation}
ds^2 = \left ( \frac{r^2}{a_0} + br + c + \frac{d}{r} \right ) dt^2 - 
\frac{dr^2}{\left ( \frac{r^2}{a_0} + br + c + \frac{d}{r} \right )}-
r^2 d\Omega ^2 ,
\label{2-1}
\end{equation}
with an electromagnetic  one-form potential:
\begin{equation}
A = A_i dx^i = A_t dt ,
\label{2-2}
\end{equation}
where $a_0,b,c,d$ and $q$ are some constants with the following 
relation (16) where $q$ is the electric charge and the parameter
 $\beta$
 used in [3] is related to our conventions via $ \beta \kappa = 4$: 
\begin{equation}
3bd - c^2 + 1 + \frac{3}{8}q^2 \kappa = 0
\label{2-3}
\end{equation}
If $b=c=d=0$ we have the 
following solution: 
\begin{equation}
ds^2 = r^2\left ( \frac{dt^2}{a_0} - \frac{a_0}{r^4}dr^2 - 
d\Omega ^2\right ) .
\label{2-4}
\end{equation}
We can introduce  new dimensionless coordinates 
$t' = t/a_0$ and $x=\sqrt {a_0}/r$ then: 
\begin{equation}
ds^2 = \frac{a_0}{x^2} \left ( dt^2 - dx^2 - d\Omega ^2\right )
\label{2-5}
\end{equation}
with $x\in(-\infty,+\infty)$ and we rename $t' \rightarrow t$. 
This metric is conformally equivalent 
to 
\begin{equation}
ds^2 = dt^2 - dx^2 - d\Omega ^2 
\label{2-6}
\end{equation}
which represents the cartesian product of a flat and a non-flat 
2-space 
of constant curvature. (For ease of comparison we mention the 
following
 result from \cite{xxx}: The cartesian product of two 2-spaces of 
constant 
curvature solves the Bach equation if and only if $X^2=Y^2$.)

Because  of the conformal invariance of the Weyl-Maxwell 
theory we can work in this paragraph with this metric (\ref{2-6}). 
We
 already  
mentioned  that this is a tube and now we can say that this tube is 
filled 
by an electric field 
$E_1 = F_{01} = q, 
\quad F_{02}=F_{03}=0$.\footnote{Remember
 that we use a  dimensionless coordinate system.} 
\par 
Now we wish to obtain this solution including a magnetic field 
and using 
the (2+2)-decomposition technique. 
It can seen that ansatz  (\ref{2-2}) can be generalized  by 
introducing 
the monopole, where $Q$ is the magnetic charge, 
$F_{23}=Q, \quad F_{12}=F_{13}=0$: 
\begin{equation}
A = \omega (r) dt - Q \cos\theta d\varphi .
\label{2-7}
\end{equation}
The Bach-Maxwell equation for metric (\ref{2-6})  looks  as: 
\begin{equation}
Y^2 - X^2 = 6\kappa \left (Q^2 + q^2 \right ) ,
\label{2-7a}
\end{equation}
here $X=0$ and $Y=2$ for the unit $S^2$-sphere. Note: As one
 knows, a Lorentz 
boost in the $x-t$-plane mixes electric and magnetic fields, but 
the term
$Q^2+q^2$ remains invariant. The Maxwell equations give us: 
\begin{equation}
\omega = q x + \omega _0 ,
\label{2-8}
\end{equation}
and finally 
\begin{equation} 
\kappa \left ( Q^2 + q^2\right ) = \frac{2}{3} .
\label{2-9}
\end{equation}
The electric and magnetic fields are: 
\begin{eqnarray}
E_x = F_{tx} & = & q ,
\label{2-10}\\
H_x = \epsilon _{x\theta\varphi} \sqrt{-g}F^{\theta\varphi} 
& = & Q .
\label{2-11}
\end{eqnarray}
here $\epsilon _{ijk}$ is the totally antisymmetric 
Levi-Civita tensor of the spatial metric. 
Thus, this (\ref{2-6}), (\ref{2-8}) and (\ref{2-9})
solution of Bach-Maxwell equations with the constant electric
 and 
magnetic fields across the area $4\pi$ of $S^2$ sphere we can 
name as {\it a flux tube}. The fluxes of electric  and 
magnetic fields across unit area of $S^2$ sphere are: 
\begin{eqnarray}
\Phi _E = 4\pi E_x & = & 4\pi q ,
\label{2-12}\\
\Phi _H = 4\pi H_x & = & 4\pi Q .
\label{2-13}
\end{eqnarray}
This is  analogous to the Levi-Civita-Robertson-Bertotti solution 
in 
 Einstein gravity \cite{levi-civita17}, \cite{robinson59}, 
\cite{bertotti59}. 
\par 
Finally, we would like to pay attention to an exceptional 
simplicity of (\ref{2-6}), (\ref{2-10}), (\ref{2-11}) 
solution with the constant electrical, magnetic fields 
and metric components.

\section{$SU(2)$ flux tube}

The spherically symmetric ansatz for an $SU(2)$ gauge field is 
a monopole-like potential written in a polar coordinate system: 
\begin{eqnarray}
A^a_t & = & v(r)\{\cos \theta ; \sin \theta \sin \varphi ; 
\sin \theta \cos \varphi \} ,
\label{3-1}\\
A^a_r & = & 0 ,
\label{3-2}\\
A^a_\theta & = & \left ( 1 -f(r) \right )
\{0; \cos \varphi ; -\sin \varphi\} ,
\label{3-3}\\
A^a_\varphi & = & \left ( 1 -f(r) \right ) \sin\theta 
\{\sin\theta; -\cos\theta \sin\varphi; -\cos\theta \cos\varphi\}
\label{3-4}
\end{eqnarray}
The Yang-Mills equations for the metric 
\begin{equation}
ds^2 = e^{2\nu}dt^2 - dr^2 - d\Omega ^2 .
\label{3-5}
\end{equation}
 are: 
\begin{eqnarray}
f'' - f'\nu ' - f\left ( f^2 - 1\right ) + e^{-2\nu}fv^2 
& = & 0 ,
\label{3-6}\\
v'' - v'\nu ' - 2vf^2 & = & 0
\label{3-7}
\end{eqnarray}
and the Bach equations are: 
\begin{eqnarray}
\frac{1}{3} X^{;A}_{;B} + \delta ^A_B 
\left ( - \frac{1}{3} \Box X + \frac{1}{12} Y^2 - 
\frac{1}{12}X^2 \right ) & = & \kappa T^A_B ,
\label{3-8}\\
\delta ^\alpha_\beta \left ( \frac{1}{6} \Box X + 
\frac{1}{12}X^2 - \frac{1}{12}Y^2\right ) & = & \kappa 
T^\alpha_\beta .
\label{3-9}
\end{eqnarray}
here $Y = 2$ is the Ricci scalar of the sphere  $S^2$.  
The energy-momentum  tensor for this $SU(2)$ gauge potential
 is: 
\begin{eqnarray}
T^t_t & = & f'^2 + \frac{1}{2}e^{-2\nu} v'^2 + 
\frac{1}{2} \left ( f^2 - 1\right )^2 + 
e^{-2\nu} f^2v^2 ,
\label{3-10}\\
T^r_r & = & -f'^2 + \frac{1}{2}e^{-2\nu} v'^2 + 
\frac{1}{2} \left ( f^2 - 1\right )^2 - 
e^{-2\nu} f^2v^2
\label{3-11}\\
T^\theta_\theta = T^\varphi_\varphi & = & 
-\frac{1}{2}e^{-2\nu} v'^2 - 
\frac{1}{2} \left ( f^2 - 1\right )^2
\label{3-12}
\end{eqnarray}
The traces of eqs. (\ref{3-8}) and (\ref{3-9})  
and the remainder of the  $\left ({t \atop t}\right )$ equation 
give us: 
\begin{eqnarray}
X^{;t}_{;t} - X^{;r}_{;r} & = & 6\kappa 
\left ( f'^2 + e^{-2\nu} f^2v^2 \right ) ,
\label{3-13}\\
\Box X + \frac{1}{2}\left (X^2 - Y^2 \right ) & = & 
-3 \kappa \left [ e^{-2\nu}v'^2 + \left ( f^2 - 1\right )^2\right ] .
\label{3-14}
\end{eqnarray}
We see that in this  case these Bach-Yang-Mills equations 
are  differential equations of 4$^{th}$ order. Only one 
way is that exclude the terms with $e^{-2\nu}$ in (\ref{3-13}) 
and (\ref{3-14}) equations: it is necessary that 
$f=0$ for (\ref{3-13}) and $e^{-2\nu}v'^2 = const$ for (\ref{3-14}). 
It is easy to show that Yang-Mills equation (\ref{3-6}) and 
(\ref{3-7}) 
have such solution: 
\begin{eqnarray}
f & = & 0 ,
\label{3-15}\\
v' & = & q e^\nu 
\label{3-16}
\end{eqnarray}
here $q$ is the some constant (color electric charge). 
In this case the simplest solution of 
(\ref{3-6}), (\ref{3-7}) (\ref{3-13}), (\ref{3-14}) is 
\begin{eqnarray}
X & = & 0 ,
\label{3-17}\\
Y & = & 2, \quad the \; curvature \; of\; S^2, 
\label{3-18}\\ 
Y^2 & = & 6\kappa \left (q^2 + 1 \right )
\label{3-19}\\
v & = & qx + v_0
\label{3-20}
\end{eqnarray}
here $v_0$ is the some constant. 

\section{Physical discussion}

It can be shown that the $SU(2)$ solution (\ref{3-15}),
 (\ref{3-16}) 
is in fact  the $U(1)$ solution. To see this most directly 
one can apply 
the following gauge transformation to the potentials of 
Eqs. (\ref{3-1} - \ref{3-4}) and taking in account 
(\ref{3-15}) and (\ref{3-20}) 
\begin{equation} 
\label{4-1} 
A'_{\mu} = S^{-1} A_{\mu} S - i (\partial _{\mu} S^{-1} )S 
\end{equation} 
where 
\begin{equation} 
\label{4-2} 
S =  \left ( 
\begin{array} {cc} \cos {\theta \over 2} & 
-e^{-i \varphi} \sin {\theta \over 2} \\ 
e^{i \varphi} \sin {\theta \over 2} & 
\cos {\theta \over 2} 
\end{array} \right ) 
\end{equation} 
with this gauge transformation we find that the gauge 
potentials become 
\begin{eqnarray} 
\label{4-3} 
A'^a _{\theta} &=& (0; 0 ; 0) ,\\ 
\label{4-4} 
A'^a _{\varphi} &=& (\cos \theta -1 ) (0 ; 0; 1) ,\\ 
\label{4-5} 
A'^a _t &=& v(r) (0; 0 ; 1) ,
\end{eqnarray} 
so that only the $\sigma ^3$ direction in isospin space 
is non-zero. The calculation of the magnetic  
and electric fields is now the same as for the $U(1)$ case 
since the ``Abelian'' gauge transformation 
given by Eqs. (\ref{4-1}) - (\ref{4-2}) 
brings us to a gauge where the non-Abelian fields take on 
an Abelian form. 
\par 
This means that in the $SU(2)$ flux tube solution happens 
the 
dynamical symmetry breakdown from $SU(2)$ gauge group 
to $U(1)$ gauge 
group: $SU(2) \rightarrow U(1)$.  Note: The same arguments
 apply also to 
 the case that we replace $SU(2)$ by $SU(n)$ for any $n>2$.
\par 
Another remarkable peculiarity for both these solutions is that 
they are the flux tubes with the constant flux of (color) electric  
and/or (color) magnetic fields across the cross section $S^2$ of 
space-like hypersurface $t=const$. 
\par 
We would like to compare these gravitational flux tube (GFT) 
with the 
flux tube in QCD. In QCD there is an assumption that the force 
lines of the $SU(3)$ gauge field between quark-antiquark pair 
are confined into a tube in the consequence of strong interaction 
of nonabelian chromodynamic field. We want 
to underline the same analogy between these two objects 
in the QCD and the Weyl conformal gravity. Of course this is not 
an exact 
correspondence as the GFT is whole spacetime but the flux tube 
in QCD exists in some external space but we repeat again that 
we say 
about some approximate correlation between these so remote 
theories. It is surprising in general that there is some 
kind of correlation between the objects in {\it classical gravity} 
and in {\it quantum nonabelian field theory}. It has to be mentioned  
that now there is a $AdS^5\times S^5$ correspondence 
\cite{hooft1}, \cite{susskind:1995vu} and \cite{thorn1},  
but for this correspondence  the existence of 
superspace is important. 
\par 
Finally we want to extract the following  physical 
pecularities of the obtained flux tube solutions in  conformal 
Weyl gravity: 
\begin{itemize}
\item
In Weyl gravity there are the GFT solutions.
\item
In $SU(2)$ GFT takes place the dynamical breakdown of gauge 
symmetry: $SU(2) \rightarrow U(1)$. 
\item 
Flux tube solutions in Weyl gravity  can be interpreted as  
some indication that this gravity theory has a classical  
confinement mechanism in the following sense: two opposite 
electrical and/or magnetic charges carried at  infinite  
distance form {\it the infinite flux tube} space filled by constant 
electric and/or magnetic fields. We underline once again 
that this confinement mechanism is a classical and not a  
quantum effect. 
\item 
There is an analogy between some objects in 
\underline{\it classical gravity} and 
\underline{\it quantum nonabelian field theory}. 
Is this analogy 
accidental or is there  some deeper connection between classical 
gravity and quantum nonabelian field theory? 
\end{itemize}

\bigskip

\section*{Acknowledgement}
Financial support from DFG and A.-v.-Humboldt-Found. 
is gratefully acknowledged.

\bibliography{john,john2}
\bibliographystyle{prsty}

\end{document}